# Information Energy Ratio of XOR Logic Gate at Mesoscopic Scale

Xiaohu Ge, Muyao Ruan, Xiaoxuan Peng, Yong Xiao, Yang Yang

**As the size of transistors approaches the mesoscopic scale, existing energy consumption analysis methods exhibit various limits, especially when being applied to describe the non-equilibrium information processing of transistors at ultra-low voltages. The stochastic thermodynamics offers a theoretic tool to analyze the energy consumption of transistor during the non-equilibrium information processing. Based on this theory, an information energy ratio of XOR gate composed of single-electron transistors is proposed at the mesoscopic scale, which can be used to quantify the exchange between the information and energy at XOR gates. Furthermore, the energy efficiency of the parity check circuit is proposed to analyze the energy consumption of digital signal processing systems. Compared with the energy efficiency of parity check circuit adopting the 7 nm semiconductor process supply voltage, simulation results show that the energy efficiency of the parity check circuit is improved by 266% when the supply voltage is chosen at a specified value.**

## I. Introduction

With the fast growing deployment of the 5th-generation (5G) mobile communication systems, the massive data need to be processed by digital signal processing circuits. The energy consumption of digital signal processing circuits is increased quickly in 5G mobile communication systems[1]. Digital signal processing circuits are composed of three types of logic gates, AND, NOT and XOR gates, all of which are made of transistors[2]. Thanks to the recent advancement of circuit technologies, the size of transistors is now at the mesoscopic scale, e.g., sub-7 nanometers(nm). However, as the sub-7nm transistor technology is approached, digital logic circuits are inevitably becoming more and more susceptible to the thermal noise due to the aggressive voltage and gate length scaling[3], especially at the mesoscopic scale. Traditional analytical methods of digital logic circuits take into account the thermal noise from phenomenological approaches. Hence, traditional analytical methods are difficult to analyze the mesoscopic scale digital circuits due to thermal fluctions of non-equilibrium information processing[4]. To overcome the limits of traditional analytical methods, the stochastic thermodynamics is introduced to analyze the non-equilibrium information processing of transistors at the mesoscopic scale[5]. Recently, the information thermodynamic theory-based energy consumption models were explored for AND and NOT gates[6]. Unfortunately, more complex logic gates such as the XOR gate are currently under-explored at the mesoscopic scale. Moreover, clarifying the transformation between the information and energy at the mesoscopic is a key and basic problem to reduce the energy consumption of logic gates, including XOR gates for digital signal circuits. Considering the complexity of non-equilibrium information processing process of transistors, it is a great challenge to reveal the information and energy coupling models of XOR gate for digital signal circuits at the mesoscopic scale.

To describe the non-equilibrium information process in digital circuits, the information thermodynamic theory has been utilized to explore the thermal, work and entropy of digital signal processing[4, 7-10]. Freitas et al. established a stochastic thermodynamic model to investigate the irreversible entropy production of nonlinear electronic circuits subject to the thermal noise[4]. The mismatch entropy production was proposed to differentiate the entropy of the actual input distribution and that of the optimal input distribution in a thermodynamic system[7]. Recently, the energy dissipation of digital circuits has been investigated by the mismatch entropy production and Landauer's principle[8]. In particular, Koski et al. characterized the thermodynamic entropy production model of single-electron transistors and derived a generalized fluctuation theorem[9]. Wimsatt et al. analyzed the physical factors that influence the energy consumption of computation systems including the computing rate, computing error rate, storage stability, circuit modularity, etc.[10].

Most existing studies on digital circuits focus on relatively simple scenarios extended from Landauer's principle. The interaction between information and energy of transistors is one of the most basic studies for the design of digital signal circuits. Motivated by the above gaps, in this article we first derive a model to quantify the information energy ratio of XOR gate at the mesoscopic scale. The key contributions of this article are briefly summarized as follows:

1. Based on the stochastic thermodynamics and information theory, the information energy ratio of XOR gate with single-electron transistors is proposed to reveal the relationship between the information capacity and energy consumption at the mesoscopic scale. An upper bound of the information energy ratio of XOR gate is derived.

2. Based on the proposed information energy ratio of XOR gate, an energy efficiency model is proposed to analyse the energy consumption of the parity check circuit.

3. Compared with the energy efficiency of parity check circuit adopting the 7 nm semiconductor process supply voltage, simulation results show that the energy efficiency of the parity check circuit is improved by 266%, when the supply voltage is set to a chosen value.

## II. Energy consumption of XOR gate

An XOR gate is composed of four single-electron NAND gates, as shown in Fig. 1a. The dynamic electron transfer process in each NAND gate is illustrated in Fig. 1b. In this way, the physical operation process of XOR gate can be abstracted as the electron transfer process. The state distribution of the NAND gate at moment $t$ can be represented using the electron number distribution of transistors, denoted as $s^t = [n_{N_1}, n_{N_2}, n_{P_1}, n_{P_2}]$ where $n_{N_1}$ and $n_{N_2}$ represent the numbers of electrons in the N-type transistors with subscript $N_1$ and $N_2$ in Fig. 1b, $n_{P_1}$ and $n_{P_2}$ represent the number of electrons in the P-type transistors with subscript $P_1$ and $P_2$ in Fig. 1b, respectively. Considering the single-electron transistor adopted in this article, the range of electron number in transistors is configured as $n_{N_1}, n_{N_2}, n_{P_1}, n_{P_2} \in \{0,1\}$.

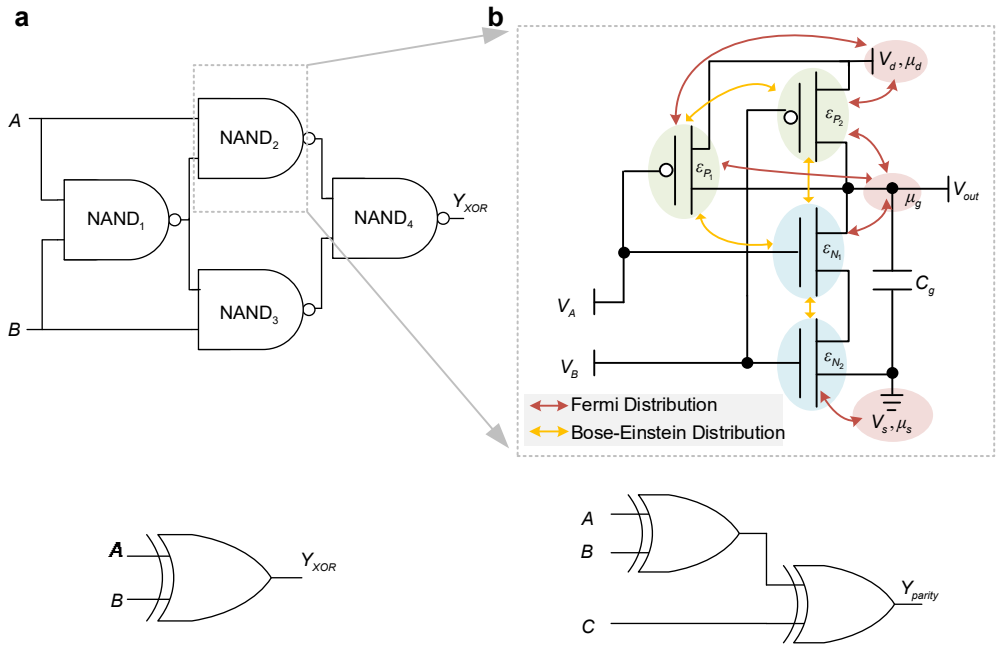

**Fig. 1 | Models of XOR gate and the parity check circuit. a**, Circuit diagram of XOR gate. The two inputs of the XOR gate are labeled as $A$ and $B$, and the output is $Y_{XOR}$. **b**, Kinetic diagram

of the gate $NAND_r$, where $r \in \{1,2,3,4\}$. Each NAND gate is composed of P-type transistors and N-type transistors. Transistors and electrodes are expressed as different energy levels $\varepsilon_j$ and chemical potentials $\mu_i$, respectively, with $j \in \{P_1, P_2, N_1, N_2\}$, $i \in \{d, s, g\}$. $V_A$ and $V_B$ are the voltages of double-inputs of the NAND gate. The output voltage of $NAND_r$ is denoted as $V_{out}$ and the load capacitance is $C_g$. The supply voltage is $V_d$, and the voltage to ground is $V_s$.

**c**, Schematic diagram of XOR gate. **d**, The parity check circuit can be composed of two XOR gates shown as XOR1 and XOR2.

When the local detailed balance condition can be satisfied[11], the transfer process of electrons within the circuit can be approximated as a Markovian process. Let $s^{t+\Delta t}$ be the state distribution of NAND gates at moment $t+\Delta t$. The state transition of NAND gate is given by

$$s^{t+\Delta t} = \mathbf{D}_{NAND} \cdot s^t, \tag{1}$$

where $\mathbf{D}_{NAND}$ is the transition rate matrix of the NAND gate, which is described as[12]

$$\mathbf{D}_{NAND} = \begin{bmatrix}
-F_1 & \delta_{dP_1}+\delta_{gP_1} & \cdots & 0 & 0 \\
\delta_{P_1 d}+\delta_{P_1 g} & -F_2 & \cdots & 0 & 0 \\
\delta_{P_2 d}+\delta_{P_2 g} & \delta_{P_2 P_1} & \cdots & 0 & 0 \\
\delta_{N_1 g} & \delta_{N_1 P_1} & \cdots & 0 & 0 \\
\delta_{N_2 s} & 0 & \cdots & 0 & 0 \\
0 & \delta_{P_2 d}+\delta_{P_2 g} & \cdots & \delta_{gN_1} & 0 \\
0 & \delta_{N_1 g} & \cdots & \delta_{dP_2}+\delta_{gP_2} & 0 \\
0 & \delta_{N_2 s} & \cdots & 0 & 0 \\
0 & 0 & \cdots & \delta_{dP_1}+\delta_{gP_1} & 0 \\
0 & 0 & \cdots & 0 & 0 \\
0 & 0 & \cdots & 0 & 0 \\
0 & 0 & \cdots & 0 & \delta_{dP_1}+\delta_{gP_1} \\
0 & 0 & \cdots & 0 & \delta_{dP_2}+\delta_{gP_2} \\
0 & 0 & \cdots & \delta_{N_2 N_1} & \delta_{gN_1} \\
0 & 0 & \cdots & -F_{14} & \delta_{sN_2} \\
0 & 0 & \cdots & \delta_{N_2 s} & -F_{15}
\end{bmatrix}, \tag{2}$$

where $F_n = \sum_{m \neq n}(\mathbf{D}_{NAND})_{m,n}$. Transfer rates between electrode $i$ and transistor $j$ are denoted as $\delta_{ji} = \Gamma f_i(\varepsilon_j)$ and $\delta_{ij} = \Gamma[1-f_i(\varepsilon_j)]$, respectively. Function $f_i(\varepsilon_j)$ is the Fermi distribution, given by $f_i(\varepsilon_j) = \left[e^{\beta(\varepsilon_j - \mu_i)} + 1\right]^{-1}$, $\Gamma$ is the rate constant, $\beta = 1/kT$, where $k$ is the Boltzmann coefficient, $T$ is the room temperature. The transfer rate between the transistors $j_1$ and $j_2$ follows the Bose-Einstein distribution, $j_1, j_2 \in \{P_1, P_2, N_1, N_2\}$.

Based on the transition rate matrix $\mathbf{D}_{\text{NAND}}$, the average electron amount $\langle n_j \rangle$ of the transistor $\varepsilon_j$ can be obtained at the present observation time interval. Furthermore, the current flowing from electrode $i$ to transistor $j$, denoted as $J_{i \to j}$, can be calculated as

$$J_{i \to j} = q\left[\delta_{ji}(1 - \langle n_j \rangle) - \delta_{ij}\langle n_j \rangle\right], \tag{3}$$

where $q$ is the unit of electric charge, $\langle n_j \rangle$ is the average number of electrons of the transistor $j$ at the present observation time. Furthermore, the energy consumption of the gate $\text{NAND}_r$ is given by

$$W_{\text{NAND}}(\tau) = \int_0^{\tau} \left[J_{s \to N_2}(\mu_s - \mu_g) + J_{d \to P_1}(\mu_d - \mu_g) + J_{d \to P_2}(\mu_d - \mu_g)\right] dt, \tag{4}$$

where the upper bound of integration $\tau$ is the propagation delay, which is taken as the moment when the difference between the output voltage and the expected voltage exceeds a certain threshold. Equation (4) shows that the computing energy consumption results from the continuous migration of electrons between transistors and electrodes.

The energy consumption of XOR gate not only depends on the input at the present moment but also the input at the previous moments[12]. Based on the four states of input at the previous moment, i.e., $\{00, 01, 10, 11\}$, the energy consumption of XOR gate based on the input state transition can be calculated as follows. Suppose a string of double-inputs sequence streams with the length $M$ is represented as $X_{in} = \{a_1b_1, a_2b_2, \ldots a_{n-1}b_{n-1}, a_nb_n, \ldots, a_Mb_M\}$, where $a_n$ and $b_n$ are the input symbols at time step $n$, $1 \leq n \leq M$, located at two inputs of XOR gate described as $A$ and $B$, respectively, with $a_nb_n \in \{00, 01, 10, 11\}$.

Considering all possible scenarios of input state transitions, the energy consumption

matrix of XOR gate can be written as

$$E_{XOR}^{diss} = \begin{pmatrix} E_{00\to 00} & E_{00\to 01} & E_{00\to 10} & E_{00\to 11} \\ E_{01\to 00} & E_{01\to 01} & E_{10\to 10} & E_{10\to 11} \\ E_{10\to 00} & E_{10\to 01} & E_{10\to 10} & E_{10\to 11} \\ E_{11\to 00} & E_{11\to 01} & E_{11\to 10} & E_{11\to 11} \end{pmatrix}, \tag{5}$$

where $E_{a_{n-1}b_{n-1}\to a_n b_n}$ is the consumed energy when the input symbol is $a_{n-1}b_{n-1}$ at the time step $n-1$ and $a_n b_n$ at the time step $n$. The input state transition matrix is

$$P_{XOR}^{trans} = \begin{pmatrix} p_{00\to 00} & p_{00\to 01} & p_{00\to 10} & p_{00\to 11} \\ p_{01\to 00} & p_{01\to 01} & p_{01\to 10} & p_{01\to 11} \\ p_{10\to 00} & p_{10\to 01} & p_{10\to 10} & p_{10\to 11} \\ p_{11\to 00} & p_{11\to 01} & p_{11\to 10} & p_{11\to 11} \end{pmatrix}, \tag{6}$$

where each element of $P_{XOR}^{trans}$ is the transition probability when the input symbol is $a_{n-1}b_{n-1}$ at the time step $n-1$ and $a_n b_n$ at the time step $n$. Based on equations (5) and (6), the average energy consumption of a single operation at XOR gate is given by

$$\overline{E}_{XOR}^{diss} = \sum_{a_{n-1}b_{n-1}} \sum_{a_n b_n} \left(E_{XOR}^{diss} \bullet P_{XOR}^{trans}\right)_{a_{n-1}b_{n-1}, a_n b_n}. \tag{7}$$

The traditional energy consumption model of XOR logic gate due to switching activities is expressed as[13]

$$P_{sw} = \zeta C V_{DD}^2 f, \tag{8}$$

where $\zeta$ is the switching coefficient, $C$ is the load capacitance, $V_{DD}$ is the supply voltage, and $f$ is the operating frequency. The average energy consumption of a single operation at XOR gate, i.e., the stochastic thermodynamic (ST) model is compared for different models in Fig. 2. When the size of transistors are assumed to be at 7 nm and simulation parameters configured as $V_d = 15 V_T$, $C_g = 1.62 \times 10^{-16}$ farad and $\alpha = 0.2$, Fig. 2a shows the energy consumption of XOR gate for a single operation with respect to the traditional model, the ST model and hierarchical simulation program for integrated circuits emphasis(HSPICE)[14]. As shown in Fig. 2a, the deviation among three models is within 10% at 7 nm CMOS technology, where the deviation of ST model is less than the deviation of traditional model in the energy consumption estimation of XOR gates. When the size of transistors is assumed to be at 0.34 nm and simulation parameters are configured as $V_d = 5V_T$, $C_g = 1.02 \times 10^{-18}$ farad and $\alpha = 0.2$. Fig. 2b shows the energy consumption of XOR gate for a single operation with respect to the traditional model, the ST model and the graphene side-wall edge gated MoS$_2$ transistor[15], where the XOR gate based on the graphene side-wall edge gated MoS$_2$ transistor consists of 16 transistors. Simulation results in Fig. 2b show that the deviation between the traditional model and graphene side-wall edge gated MoS$_2$ transistor is 73.47%. Therefore, the traditional model is no longer appropriate for analyzing the energy consumption of XOR gates. Simulation results in Fig 2b validate that the deviation between the ST model and graphene side-wall edge gated MoS$_2$ transistor is 5.92%. Hence, the ST model can still be used for analyzing the energy consumption of XOR gates.

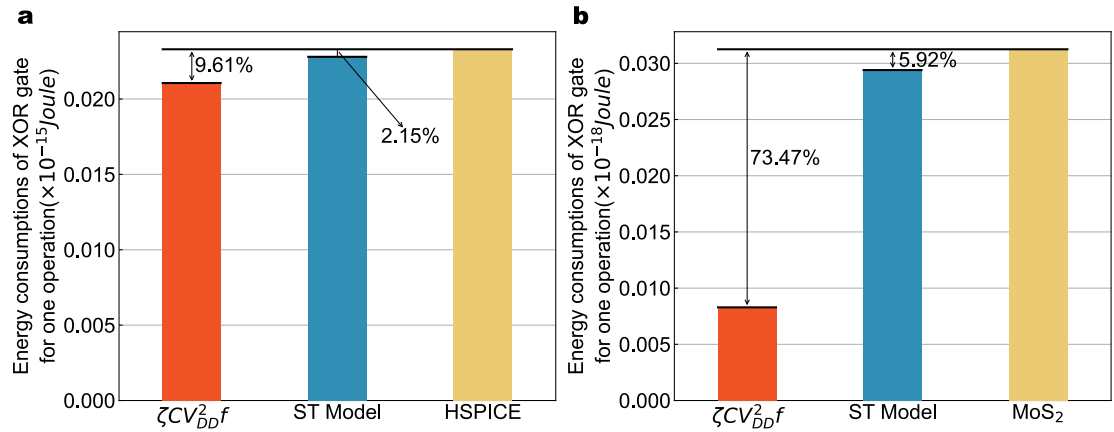

**Fig. 2 | Energy consumption of XOR gate of a single operation. a**, Energy consumption of XOR gate with respect to $\zeta CV_{DD}^2 f$, the ST model and the simulation results based on HSPICE. **b**, Energy consumption of XOR gate with respect to $\zeta CV_{DD}^2 f$, the ST model and the 0.34nm graphene side-wall edge gated MoS$_2$ transistor.

### III. Information capacity of XOR gate

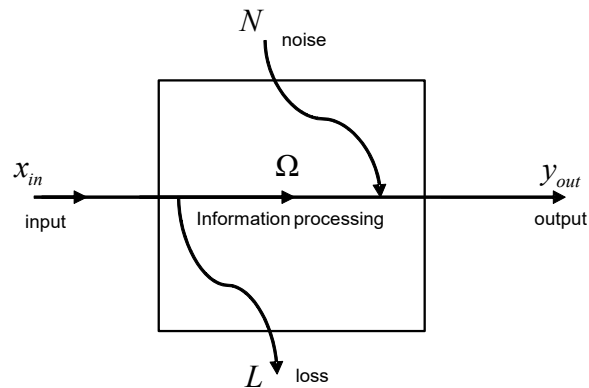

**Fig. 3 | Schematic diagram of the information processing process**

Fig. 3 illustrates the information processing process, wherein the information processing module can be an algorithm, an integrated circuit or a simple logic gate. In Fig. 3, the input symbol is $x_{in}$ and it belongs to $\mathcal{X} = \{x_1, x_2, ... x_{in}, ..., x_K\}$; the output symbol is $y_{out}$, and it belongs to $\mathcal{Y} = \{y_1, y_2, ..., ... y_{out}, ..., y_K\}$.

Let us now extend the model illustrated in Fig. 3 to XOR gate. As shown in Fig. 1c, $A$ and $B$ are the inputs of XOR gate, and $Y_{XOR}$ is the output. Based on the threshold voltage judgment criterion, the logic states of the input and output are mapped as:

$$A = \begin{cases} 0, & V_{in} = 0 \\ 1, & V_{in} = V_d \end{cases} \quad B = \begin{cases} 0, & V_{in} = 0 \\ 1, & V_{in} = V_d \end{cases} \quad Y_{XOR} = \begin{cases} 0, & V_{out} \leq (1-\alpha)V_d \\ 1, & V_{out} \geq \alpha V_d \\ \varnothing, & \text{otherwise} \end{cases}, \quad (9)$$

where $V_{in}$ is the voltage of each input, $V_{out}$ is the voltage of output. $\alpha$ is the threshold factor. When the output voltage is higher than $\alpha V_d$, the output symbol is considered as logic 1, and considered as logic 0 when $V_{out}$ is lower than $(1-\alpha)V_d$, otherwise as $\varnothing$.

Based on the stochastic thermodynamic model of XOR gate, the output voltage distribution of XOR gate is governed by Gaussian distributions based on Gillespie's algorithm[16]. Assume that the output symbol of XOR gate is $y_n^{XOR}$ at the time step $n$. The output voltage of XOR gate $V_{out}^{XOR}$ is governed by the following distribution:

$$V_{out}^{XOR} \sim \begin{cases} \mathrm{N}(0, 1/\beta C_g) & a_n \oplus b_n = 0 \\ \mathrm{N}(V_d, 1/\beta C_g) & a_n \oplus b_n = 1 \end{cases}, \quad (10)$$

where $\mathrm{N}(\cdot)$ denotes Gaussian distribution. The symbol $\oplus$ denotes the operation of XOR. Therefore, the mutual information between the input and output of XOR gate can be written as

$$I(AB;Y_{XOR}) = \sum_{a_n \in A} \sum_{b_n \in B} \sum_{y_n^{XOR} \in Y_{XOR}} p(a_n, b_n, y_n^{XOR}) \log_2 \left( \frac{p(a_n, b_n, y_n^{XOR})}{p(a_n b_n) p(y_n^{XOR})} \right)$$
$$= \sum_{a_n \in A} \sum_{b_n \in B} \sum_{y_n^{XOR} \in Y_{XOR}} p(a_n, b_n, y_n^{XOR}) \log_2 \left( \frac{p(y_n^{XOR} | a_n b_n)}{p(y_n^{XOR})} \right), \quad (11)$$

where $p(a_n, b_n, y_n^{XOR})$ is the joint probability when the input of XOR gate is $a_n b_n$ and the output symbol is $y_n^{XOR}$, $p(a_n b_n)$ is the probability when the input is $a_n b_n$, $p(y_n^{XOR} | a_n b_n)$ is the conditional probability when the output symbol is $y_n^{XOR}$ and the input symbol is $a_n b_n$, and $p(y_n^{XOR})$ is the probability when the output symbol is $y_n^{XOR}$. By iterating through all the possible input and output scenarios, the information capacity of XOR gate for a single operation is given by

$$C = \max I(AB;Y_{XOR})$$
$$= \max \sum_{a_n \in A} \sum_{b_n \in B} \sum_{y_n^{XOR} \in Y_{XOR}} p(a_n, b_n, y_n^{XOR}) \log_2 \left( \frac{p(y_n^{XOR} | a_n b_n)}{p(y_n^{XOR})} \right). \quad (12)$$

Equation (12) indicates that the maximum information rate can be achieved by transmitting information for an arbitrarily small error rate. See Appendix A for information capacity details.

### IV. Information energy ratio of XOR gate

To reveal the relationship between the information capacity and energy consumption of XOR gate at the mesoscopic scale, an information energy ratio of XOR gate is given by

$$\eta_{XOR} = \frac{I(AB;Y_{XOR})}{\overline{E}_{XOR}^{diss}} \quad (bits / kT), \quad (13)$$

where $I(AB;Y_{XOR})$ is the mutual information between the input and output of XOR gate for a single operation, $\overline{E}_{XOR}^{diss}$ is the average energy consumption of XOR gate for a single operation. $\eta_{XOR}$ indicates the amount of information transmitted while an energy of $kT$ is cost for an XOR gate. From equation (11), the mutual information of

XOR gate is derived as

$$I(AB;Y_{XOR}) = \sum_{a_n \in A} \sum_{b_n \in B} \sum_{y_n^{XOR} \in Y_{XOR}} p(a_n, b_n, y_n^{XOR}) \log_2 \left( \frac{p(y_n^{XOR} | a_n b_n)}{p(y_n^{XOR})} \right)$$

$$= (p_{00} + p_{11}) \sum_{y_n^{XOR} \in Y_{XOR}} \left[ p(y_n^{XOR} | a_n b_n = 00) \log_2 \left( \frac{p(y_n^{XOR} | a_n b_n = 00)}{p(y_n^{XOR})} \right) \right] + (p_{01} + p_{10}) \sum_{y_n^{XOR} \in Y_{XOR}} \left[ p(y_n^{XOR} | a_n b_n = 01) \log_2 \left( \frac{p(y_n^{XOR} | a_n b_n = 01)}{p(y_n^{XOR})} \right) \right]$$

$$= \left( \left\| P_{XOR}^{trans} \cdot [1 \ 0 \ 0 \ 0]^T \right\|_1 + \left\| P_{XOR}^{trans} \cdot [0 \ 0 \ 0 \ 1]^T \right\|_1 \right) \sum_{y_n^{XOR} \in Y_{XOR}} \left[ p(y_n^{XOR} | a_n b_n = 00) \log_2 \left( \frac{p(y_n^{XOR} | a_n b_n = 00)}{p(y_n^{XOR})} \right) \right] +$$

$$\left( \left\| P_{XOR}^{trans} \cdot [0 \ 1 \ 0 \ 0]^T \right\|_1 + \left\| P_{XOR}^{trans} \cdot [0 \ 0 \ 1 \ 0]^T \right\|_1 \right) \sum_{y_n^{XOR} \in Y_{XOR}} \left[ p(y_n^{XOR} | a_n b_n = 01) \log_2 \left( \frac{p(y_n^{XOR} | a_n b_n = 01)}{p(y_n^{XOR})} \right) \right],$$

(14)

where $p_{ab}$ is the probability when the input symbols of XOR gate are $a_n$ and $b_n$ at the time step $n$. The norm of vector $l$ is $\|l\|_1 = \sum_{i=1}^{n} |l_i|$. Superscript T is transposing the matrix. Combined with equations (7) and (14), equation (13) can be rewritten as

$$\eta_{XOR} = \frac{\begin{bmatrix} \left( \left\| P_{XOR}^{trans} \cdot [1 \ 0 \ 0 \ 0]^T \right\|_1 + \left\| P_{XOR}^{trans} \cdot [0 \ 0 \ 0 \ 1]^T \right\|_1 \right) \sum_{y_n^{XOR} \in Y_{XOR}} \left[ p(y_n^{XOR} | a_n b_n = 00) \log_2 \left( \frac{p(y_n^{XOR} | a_n b_n = 00)}{p(y_{XOR})} \right) \right] + \\ \left( \left\| P_{XOR}^{trans} \cdot [0 \ 1 \ 0 \ 0]^T \right\|_1 + \left\| P_{XOR}^{trans} \cdot [0 \ 0 \ 1 \ 0]^T \right\|_1 \right) \sum_{y_n^{XOR} \in Y_{XOR}} \left[ p(y_n^{XOR} | a_n b_n = 01) \log_2 \left( \frac{p(y_n^{XOR} | a_n b_n = 01)}{p(y_n^{XOR})} \right) \right] \end{bmatrix}}{\sum_{a_{n-1} b_{n-1}} \sum_{a_n b_n} \begin{pmatrix} E_{00 \to 00} p_{00 \to 00} & E_{00 \to 01} p_{00 \to 01} & E_{00 \to 10} p_{00 \to 10} & E_{00 \to 11} p_{00 \to 11} \\ E_{01 \to 00} p_{01 \to 00} & E_{01 \to 01} p_{01 \to 01} & E_{01 \to 10} p_{01 \to 10} & E_{01 \to 11} p_{01 \to 11} \\ E_{10 \to 00} p_{10 \to 00} & E_{10 \to 01} p_{10 \to 01} & E_{10 \to 10} p_{10 \to 10} & E_{10 \to 11} p_{10 \to 11} \\ E_{11 \to 00} p_{11 \to 00} & E_{11 \to 01} p_{11 \to 01} & E_{11 \to 10} p_{11 \to 10} & E_{11 \to 11} p_{11 \to 11} \end{pmatrix}_{a_{n-1} b_{n-1}, a_n b_n}}.$$

(15)

The optimization problem to maximize the information energy ratio of XOR gate can be written as

$$\begin{aligned} \max \eta_{XOR} &= f(V_d, p_a, p_b) \\ \text{s.t.} \quad 4V_T &\leq V_d \leq 6V_T, \ 0 \leq p_a \leq 1, \ 0 \leq p_b \leq 1 \end{aligned}, \quad (16)$$

where $f(\cdot)$ is the function operation based on the equation (14), $p_a$ and $p_b$ are probabilities when the input symbol is logic 0 at the double-inputs of XOR, respectively, $V_T = kT/q$ is the thermal voltage. Since the equation (14) is a discontinuous function, the Genetic Algorithm[18] (GA) is adopted to obtain the upper bound of the information energy ratio of XOR gate. In GA the binary coding is used to encode individuals, here each individual can be regarded as a solution to the above problem (16) and each individual owns a set of chromosomes represented by three variables $V_d$, $p_a$ and $p_b$.

We consider the following simulation configuration: the length of gene code is 10, the size of the population is 80, the evolution times are 100, the mutation probability is 0.001. The maximum of information energy ratio is configured as the fitness function of GA. The upper bound of information energy ratio is obtained for XOR gate during continuous iterations. Based on the simulation results, we can observe that the upper bound of information energy ratio of XOR gate converges to 0.00026 bits/kT, i.e., the red line in Fig. 4, which is approximated when $V_d \approx 5.16 V_T$, $p_a \approx 0.39$ and $p_b \approx 1$.

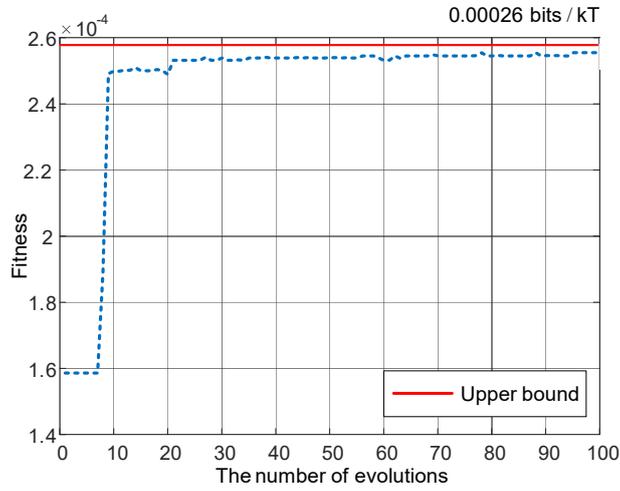

**Fig. 4 | Energy efficiency of XOR gate under different numbers of evolutions. The blue dashed line shows the information energy ratio of XOR gate converges as the number of evolutions increases, and the red solid line is the upper bound of information energy ratio for XOR gate.**

## V. Energy efficiency of parity check circuit

Based on the information energy ratio of XOR gate, let us now investigate the energy efficiency of parity check circuit. Fig. 1d is a schematic diagram of parity check circuit which is composed of two XOR gates connected in series, with the three inputs of parity check circuit represented as $A$, $B$ and $C$, and the output represented as $Y_{parity}$, $A, B, C \in \{0,1\}$, $Y_{parity} \in Y_{XOR}$. Parity check circuit performs the function of detecting the number of logic 1 among three input symbols of the circuit.

The sequences of three input symbols with length $M$ are represented as $S_A = \{a_1, a_2, ... a_{n-1}, a_n, ..., a_M\}$, $S_B = \{b_1, b_2, ... b_{n-1}, b_n, ..., b_M\}$ and

$S_C = \{c_1, c_2, \ldots c_{n-1}, c_n, \ldots, c_M\}$, through the input $A$, $B$ and $C$. $a_n$, $b_n$ and $c_n$ represent the three input symbols at the time step $n$ with $a_n b_n c_n \in \{000, 001, 010, 011, 100, 101, 110, 111\}$. Suppose that the three inputs of parity check circuit are independent identity distributions (i.i.d.). Based on the mutual information of XOR gate, the mutual information at the input and output of the parity check circuit at the time step $n$ is

$$I(ABC; Y_{parity}) = \sum_{a_n \in A} \sum_{b_n \in B} \sum_{c_n \in C} \sum_{y_n^{parity} \in Y_{XOR}} p(a_n, b_n, c_n, y_n^{parity}) \log_2 \left( \frac{p(a_n, b_n, c_n, y_n^{parity})}{p(a_n b_n c_n) p(y_n^{parity})} \right)$$
$$= \sum_{a_n \in A} \sum_{b_n \in B} \sum_{c_n \in C} \sum_{y_n^{parity} \in Y_{XOR}} p(a_n, b_n, c_n, y_n^{parity}) \log_2 \left( \frac{p(y_n^{parity} | a_n b_n c_n)}{p(y_n^{parity})} \right).$$

(17)

where $y_n^{parity}$ is the output symbol of XOR2 at the time step $n$ in Fig. 1d, $y_n^{parity} \in Y_{parity}$. $p(a_n, b_n, c_n, y_n^{parity})$ is the joint probability when the three input symbols of the parity check circuit are $a_n b_n c_n$ and the output symbol is $y_n^{parity}$, $p(a_n b_n c_n)$ is the probability when the three input symbols are $a_n b_n c_n$, $p(y_n^{parity} | a_n b_n c_n)$ is the conditional probability that the output symbol is $y_n^{parity}$ when the three input symbols are $a_n b_n c_n$, $p(y_n^{parity})$ is the probability that the output symbol is $y_n^{parity}$. For all eight cases of the inputs and three cases of the output, the mutual information between the input and output of parity check circuit can be obtained by traversing all 24 cases.

The average energy consumption of a single computing of parity check circuit is expressed as

$$\overline{E}_{parity}^{diss} = \sum_{a_{n-1} b_{n-1} c_{n-1}} \sum_{a_n b_n c_n} \left( E_{XOR}^{diss} \bullet P_{XOR1}^{trans} + E_{XOR}^{diss} \bullet P_{XOR2}^{trans} \right)_{a_{n-1} b_{n-1} c_{n-1}, a_n b_n c_n}, \quad (18)$$

where $P_{XOR1}^{trans}$ and $P_{XOR2}^{trans}$ are the input state transition matrix of XOR1 and XOR2, respectively. Derivation of equation (18) is given in Appendix B.

Combining with equations (13), (17) and (18), the energy efficiency of parity check circuit is

$$\eta_{parity} = I(ABC; Y_{parity}) / \overline{E}_{parity}^{diss}$$

$$= \frac{\sum_{a_n \in A} \sum_{b_n \in B} \sum_{c_n \in C} \sum_{y_n^{parity} \in Y_{XOR}} p(a_n, b_n, c_n, y_n^{parity}) \log_2 \left( \frac{p(y_n^{parity} | a_n b_n c_n)}{p(y_n^{parity})} \right)}{\sum_{a_{n-1} b_{n-1} c_{n-1}} \sum_{a_n b_n c_n} \left( E_{XOR}^{diss} \cdot P_{XOR1}^{trans} + E_{XOR}^{diss} \cdot P_{XOR2}^{trans} \right)_{a_{n-1} b_{n-1} c_{n-1}, a_n b_n c_n}}. \quad (19)$$

The supply voltage of XOR gate is the key factors to determine the error probability of XOR gate. Moreover, the energy efficiency of parity check circuit depends on the optimization of supply voltage. Without loss of generality, when the previous input is $a_{n-1} b_{n-1} = 00$, the error probability of XOR gate with respect to the supply voltage is presented in Fig. 5 which shows that the error probability of XOR gate is larger than 1% when the supply voltage is less than $4V_T$. When the supply voltage is larger than or equal to $4V_T$, the error probability of XOR gate is less than 1%. Therefore, the XOR gate can function well only the supply voltage is larger than or equal to $4V_T$.

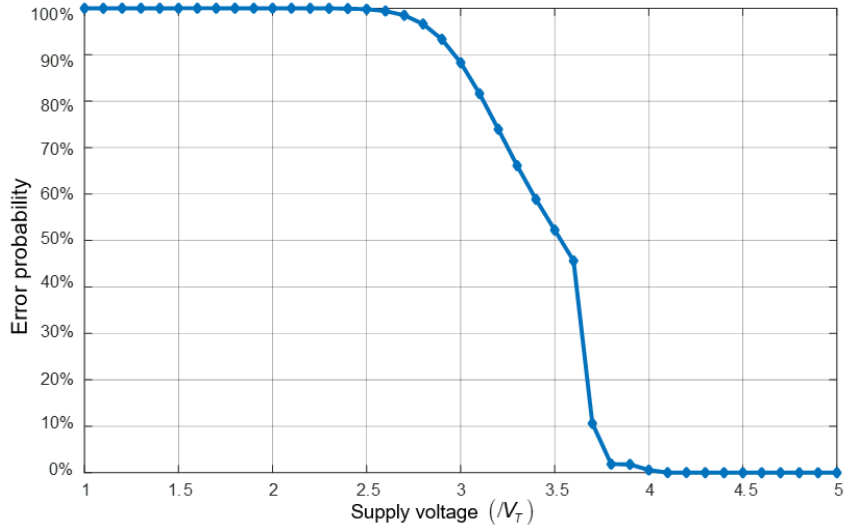

**Fig. 5 | Error probability of XOR gate with respect to the supply voltage when the previous input is** $a_{n-1} b_{n-1} = 00$.

In Fig. 6a, an image pixel processing scene is considered as a case study for analyzing the energy efficiency of parity check circuit. Based on the results in Fig. 5, the supply voltage of XOR gate is configured to be larger than or equal to $4V_T$. When the size of transistors is manufactured at 7 nm, the supply voltage of XOR gate is 0.39 Volt[19], i.e., $15V_T$. Compared with the energy efficiency of the parity check circuit with 7 nm semiconductor process supply voltage, i.e., $15V_T$, Fig. 6b shows that the energy efficiency of parity check circuit with the supply voltage $5V_T$ is improved by 266%.

## VI. Conclusions

A stochastic thermodynamic theory-based energy consumption model is proposed for XOR gates. Based on information theory, the mutual information of XOR gate is derived for information processing. Furthermore, the information energy ratio of XOR gate is established to reveal the relationship between the information capacity and energy consumption at the mesoscopic scale. Based on the information energy ratio, the energy efficiency model of one of typical digital signal processing circuits, i.e., the parity check circuit is established. Compared with the energy efficiency of parity check circuit adopting the normal 7 nm semiconductor process supply voltage, i.e., $15V_T$, simulation results show that the energy efficiency of parity check circuit with the supply voltage $5V_T$ can be improved by 266%. In future work, we plan to explore new guidelines to improve the energy efficiency of complex digital signal processing circuits.

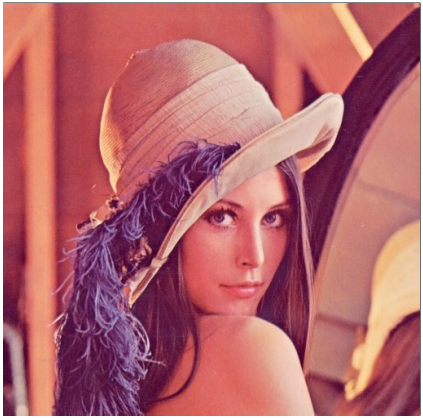 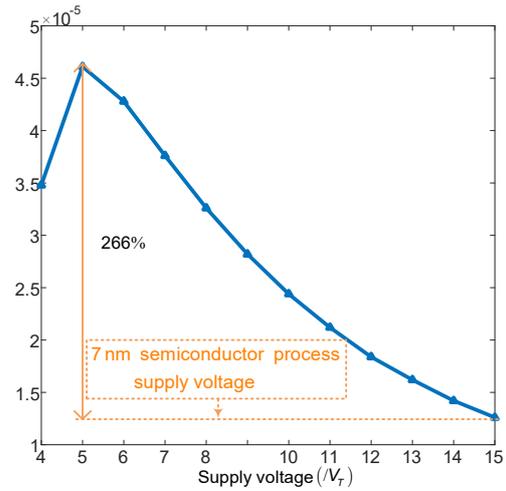

**Fig. 6 | Simulation results of the energy efficiency of the parity check circuit. a,** Example of the selected pictures. **b,** Energy efficiency with respect to the supply voltages.

## VII.    Methods

All of our results were achieved through simulations. The properties of the distribution of the output voltage are observed by both an iterative, numerically diagonalization of the master equation and the Gillespie algorithm realized by code. With the simulation results which show the voltage of output is governed by a Gaussian distribution, the mutual information of XOR gate is given based on the information theory, as shown in equation (11). Genetic Algorithm is utilized to obtain the upper bound of the information energy ratio of XOR gate. All our code is performed on PyCharm and MATLAB.

## VIII. Supplementary information

**Appendix A: Information capacity details**

In this appendix, the details about information capapcity of XOR is given. Based on Fig. 2, the amount of information at the input and output are expressed as follows:

$$I_{in} = \sum_{x_{in} \in \mathcal{X}} p(x_{in}) \log_2 \left( \frac{1}{p(x_{in})} \right), \tag{A1}$$

$$I_{out} = \sum_{y_{out} \in \mathcal{Y}} p(y_{out}) \log_2 \left( \frac{1}{p(y_{out})} \right), \tag{A2}$$

where $p(x_{in})$ is the probability when the input symbol is $x_{in}$, $p(y_{out})$ is the probability when the output symbol is $y_{out}$.

Considering all possible input states, we denote the noise $N$ of a process as the uncertainty in the output given the input state:

$$N = \sum_{x_{in} \in \mathcal{X}} p(x_{in}) \sum_{y_{out} \in \mathcal{Y}} p(y_{out} | x_{in}) \log_2 \left( \frac{1}{p(y_{out} | x_{in})} \right), \tag{A3}$$

where $p(y_{out} | x_{in})$ is the conditional probability of output $y_{out}$ given input $x_{in}$. The mutual information between input and output is expressed as $\Omega$:

$$\Omega = I_{out} - N. \tag{A4}$$

Based on equation (A4), the mutual information in equation (11) is obtained. The maximum of mutual information is the information capacity of XOR gate in equation (12).

The input states of XOR gate are divided into 4 cases: $p(y_n^{XOR} = 0 | a_n b_n \in \{00,11\})$, $p(y_n^{XOR} = 1 | a_n b_n \in \{01,10\})$, $p(y_n^{XOR} = 0 | a_n b_n \in \{01,10\})$ and $p(y_n^{XOR} = 1 | a_n b_n \in \{00,11\})$. When $a_n b_n \in \{00,11\}$, the error probability is expressed as

$$\xi_{a_n b_n \in \{00,11\}} = p(y_n^{XOR} \neq 0 | a_n b_n \in \{00,11\}) = \Phi\left(\frac{a_n b_n}{\sigma}\right) \approx 1 - \frac{1}{2\sqrt{\pi \beta C_g}} \frac{e^{-\beta C_g (1-\alpha)^2 V_d^2}}{(1-\alpha)V_d}, \tag{A5}$$

where the function $\Phi$ is the cumulative normal distribution function of the standard Gaussian distribution, given by $\Phi(\omega) = \frac{1}{\sqrt{2\pi}} \int_{-\infty}^{\omega} e^{-\theta^2/2} d\theta$, where $\omega = a_n b_n / \sigma$, $\sigma = \sqrt{1/\beta C_g}$. When $a_n b_n \in \{01,10\}$, the error probability is expressed as

$$\xi_{a_n b_n \in \{01,10\}} = p(y_n^{XOR} \neq 1 | a_n b_n \in \{01,10\}) = \Phi\left(\frac{a_n b_n - V_d}{\sigma}\right) = \frac{1}{2\sqrt{\pi \beta C_g}} \frac{e^{-\beta C_g \alpha^2 V_d^2}}{\alpha V_d}.$$

(A6)

Based on equations (A5) and (A6), the transition probability of XOR gate in equation (12) is derived as

$$\begin{cases} p(y_n^{XOR} = 0 | a_n b_n \in \{01,10\}) = \xi_{a_n b_n \in \{00,11\}} \\ p(y_n^{XOR} = 1 | a_n b_n \in \{01,10\}) = 1 - \xi_{a_n b_n \in \{00,11\}} \\ p(y_n^{XOR} = 0 | a_n b_n \in \{00,11\}) = 1 - \xi_{a_n b_n \in \{01,10\}} \\ p(y_n^{XOR} = 1 | a_n b_n \in \{00,11\}) = \xi_{a_n b_n \in \{01,10\}} \end{cases}. \quad (A7)$$

**Appendix B: Detailed energy consumption derivation of parity check circuit**

$P_{XOR1}^{trans}$ and $P_{XOR2}^{trans}$ are the input state transition matrixs of XOR1 and XOR2 in equation (18), which are denoted as:

$$P_{XOR1}^{trans} = \begin{pmatrix} p_{a_n}^2 p_{b_n}^2 & p_{a_n}^2 p_{b_n} \overline{p_{b_n}} & p_{a_n} \overline{p_{a_n}} p_{b_n}^2 & p_{a_n} \overline{p_{a_n}} p_{b_n} \overline{p_{b_n}} \\ p_{a_n}^2 p_{b_n} \overline{p_{b_n}} & p_{a_n}^2 \overline{p_{b_n}}^2 & p_{a_n} \overline{p_{a_n}} p_{b_n} \overline{p_{b_n}} & p_{a_n} \overline{p_{a_n}} \overline{p_{b_n}}^2 \\ p_{a_n} \overline{p_{a_n}} p_{b_n}^2 & p_{a_n} \overline{p_{a_n}} p_{b_n} \overline{p_{b_n}} & \overline{p_{a_n}}^2 p_{b_n}^2 & \overline{p_{a_n}}^2 p_{b_n} \overline{p_{b_n}} \\ p_{a_n} \overline{p_{a_n}} p_{b_n} \overline{p_{b_n}} & p_{a_n} \overline{p_{a_n}} \overline{p_{b_n}}^2 & \overline{p_{a_n}}^2 p_{b_n} \overline{p_{b_n}} & \overline{p_{a_n}}^2 \overline{p_{b_n}}^2 \end{pmatrix}, \quad (B1)$$

$$P_{XOR2}^{trans} = \begin{pmatrix} \lambda_1 p_{c_n}^2 & \lambda_1 p_{c_n} \overline{p_{c_n}} & \lambda_2 p_{c_n}^2 & \lambda_2 p_{c_n} \overline{p_{c_n}} \\ \lambda_1 p_{c_n} \overline{p_{c_n}} & \lambda_1 \overline{p_{c_n}}^2 & \lambda_2 p_{c_n} \overline{p_{c_n}} & \lambda_2 \overline{p_{c_n}}^2 \\ \lambda_3 p_{c_n}^2 & \lambda_3 p_{c_n} \overline{p_{c_n}} & \lambda_4 p_{c_n}^2 & \lambda_4 p_{c_n} \overline{p_{c_n}} \\ \lambda_3 p_{c_n} \overline{p_{c_n}} & \lambda_3 \overline{p_{c_n}}^2 & \lambda_4 p_{c_n} \overline{p_{c_n}} & \lambda_4 \overline{p_{c_n}}^2 \end{pmatrix}, \quad (B2)$$

where each matrix element of $P_{XOR1}^{trans}$ and $P_{XOR2}^{trans}$ is the transition probability of the two input symbols of XOR1 and XOR2 between the time step $n-1$ and $n$, respectively.

For simplicity, $\lambda_1$, $\lambda_2$ and $\lambda_3$ are introduced and are expressed as follows:

$$\begin{cases} \lambda_1 = p_{a_n}^2 p_{b_n}^2 + 2 p_{a_n} \overline{p_{a_n}} p_{b_n} \overline{p_{b_n}} + \overline{p_{a_n}}^2 \overline{p_{b_n}}^2 \\ \lambda_2 = p_{a_n}^2 p_{b_n} \overline{p_{b_n}} + p_{a_n} \overline{p_{a_n}} p_{b_n}^2 + p_{a_n} \overline{p_{a_n}} \overline{p_{b_n}}^2 + \overline{p_{a_n}}^2 p_{b_n} \overline{p_{b_n}} \\ \lambda_3 = p_{a_n}^2 p_{b_n} \overline{p_{b_n}} + p_{a_n} \overline{p_{b_n}} p_{b_n}^2 + p_{a_n} \overline{p_{a_n}} \overline{p_{b_n}}^2 + \overline{p_{a_n}}^2 p_{b_n} \overline{p_{b_n}} \end{cases}, \qquad (B3)$$

where $\overline{p_s} = 1 - p_s$, $s \in \{a_n, b_n, c_n\}$. $p_{a_n}$ is the probability when the input symbol $a_n$ is logic 0, $p_{b_n}$ is the probability when the input symbol $b_n$ is logic 0, and $p_{c_n}$ is the probability when the input symbol $c_n$ is logic 0.

Furthermore, in equation (18) which gives the average energy consumption of a single computing of parity check circuit, $\left( E_{XOR}^{diss} \cdot P_{XOR1}^{trans} + E_{XOR}^{diss} \cdot P_{XOR2}^{trans} \right)$ is expressed as:

$$\begin{cases} E_{00\to 00}\left(p_{a_n}^2 p_{b_n}^2 + \lambda_1 p_{c_n}^2\right) & E_{00\to 01}\left(p_{a_n}^2 p_{b_n} \overline{p_{b_n}} + \lambda_1 p_{c_n} \overline{p_{c_n}}\right) & E_{00\to 10}\left(p_{a_n} \overline{p_{a_n}} p_{b_n}^2 + \lambda_2 p_{c_n}^2\right) & E_{00\to 11}\left(p_{a_n} \overline{p_{a_n}} p_{b_n} \overline{p_{b_n}} + \lambda_2 p_{c_n} \overline{p_{c_n}}\right) \\ E_{01\to 00}\left(p_{a_n}^2 p_{b_n} \overline{p_{b_n}} + \lambda_1 p_{c_n} \overline{p_{c_n}}\right) & E_{01\to 01}\left(p_{a_n}^2 \overline{p_{b_n}}^2 + \lambda_1 \overline{p_{c_n}}^2\right) & E_{10\to 10}\left(p_{a_n} \overline{p_{a_n}} p_{b_n} \overline{p_{b_n}} + \lambda_2 p_{c_n} \overline{p_{c_n}}\right) & E_{10\to 11}\left(p_{a_n} \overline{p_{a_n}} \overline{p_{b_n}}^2 + \lambda_2 \overline{p_{c_n}}^2\right) \\ E_{10\to 00}\left(p_{a_n} \overline{p_{a_n}} p_{b_n}^2 + \lambda_3 p_{c_n}^2\right) & E_{10\to 01}\left(p_{a_n} \overline{p_{a_n}} p_{b_n} \overline{p_{b_n}} + \lambda_3 p_{c_n} \overline{p_{c_n}}\right) & E_{10\to 10}\left(\overline{p_{a_n}}^2 p_{b_n}^2 + \lambda_4 p_{c_n} \overline{p_{c_n}}\right) & E_{10\to 11}\left(\overline{p_{a_n}}^2 p_{b_n} \overline{p_{b_n}} + \lambda_4 p_{c_n} \overline{p_{c_n}}\right) \\ E_{11\to 00}\left(p_{a_n} \overline{p_{a_n}} p_{b_n} \overline{p_{b_n}} + \lambda_3 p_{c_n} \overline{p_{c_n}}\right) & E_{11\to 01}\left(p_{a_n} \overline{p_{a_n}} \overline{p_{b_n}}^2 + \lambda_3 \overline{p_{c_n}}^2\right) & E_{11\to 10}\left(\overline{p_{a_n}}^2 p_{b_n} \overline{p_{b_n}} + \lambda_4 p_{c_n} \overline{p_{c_n}}\right) & E_{11\to 11}\left(\overline{p_{a_n}}^2 \overline{p_{b_n}}^2 + \lambda_4 \overline{p_{c_n}}^2\right) \end{cases}.$$

(B4)